# RESPONSE TO FENG HAO'S PAPER "KISH'S KEY EXCHANGE SCHEME IS INSECURE" [*]

L. B. KISH[†],

*Department of Electrical and Computer Engineering, Texas A&M University, College Station, TX 77843-3128, USA*

In a forthcoming paper in *IEE Proceedings Information Security*, Feng Hao claims that temperature inaccuracies make the key exchange scheme based on *Kirchhoff Loop with Johnson-like Noise* insecure. First we point out that this claim is irrelevant for the security of the idealized/mathematical communicator because it violates basic assumptions. Furthermore, in general practical applications, as it has been pointed out in the very first paper, the use of thermal noise is disadvantageous therefore the issue of temperature and its accuracy is unimportant even for the security of common practical realizations. It is important to emphasize that any deviation from the idealized scheme can lead to a finite level of security. Thus, if the above-mentioned author had stressed the inaccuracy of the resistor values; his claim would have been practically valid. However the impact of such systematic errors can be kept under control. Here we cite our relevant analysis (*Phys. Lett. A* **359**, (2006) 741-744) of 1% voltage inaccuracy along the line due to wire resistance effects, which can be considered as a general result for inaccuracies of this order. The necessity to build a statistics to extract information during the properly chosen clock time leads to extremely poor eavesdropper statistics and this situation provides a good practical raw bit security that can be orders of magnitude beyond the raw bit security of idealized quantum communicators.

## 1. Feng Hao's comments

In a forthcoming paper [1], Feng Hao claims that temperature inaccuracy/difference at the resistors in the *Kirchhoff Loop Johnson-like Noise* (KLJN) communicators [2-8] compromises its security. Indeed, if we run the communicator with itself the Johnson noise of the resistors (stealth communication [9]) and the temperature of one of the resistors deviates from the temperature of the other resistors, there will be a net power flow between the communicators (Alice and Bob), see Figure 1. As it was pointed out in the first paper [2], non-zero net power flow between Alice and Bob compromises security. In the present case, if the clock period were infinite, the situation when the resistor with different temperature is connected to the line could be identified by the measurement of the net power flow and its direction. Though the eavesdropper (Eve) cannot identify the bit, high or low, she can identify this situation whenever it happens, provided the temperature is fixed and the clock period is long enough. That means 50% of the shared key or its inverse could be extracted. Similarly, with proper differences between the temperatures of the other resistances, the whole secure key or its inverse could be extracted by Eve provided the clock period is sufficiently long. After getting the key or its inverse, Eve can test it and its inverse on the message and can decrypt the data.

---

[*] In an email on October 29, 2006, the author of this paper asked the Editors in Chief of the *IEE Proceedings Information Security* for an opportunity of publishing a response following directly Feng Hao's paper. However the email answer (November 20, 2006) contained a negative response saying that such practice "is not the custom in respected venues in cryptography".

[†] Until 1999, L.B. Kiss

*Response to Feng Hao's comments about security versus temperature inaccuracies.*

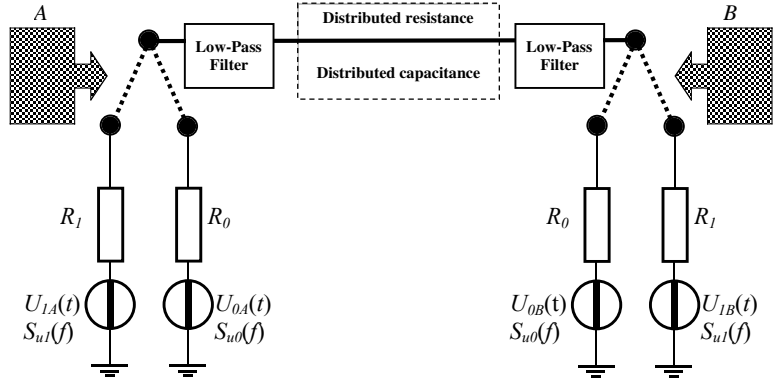

**Figure 1.** The Kirchhoff Loop Johnson-like Noise (KLJN) setup. Feng Hao's comment is based on the assumption that the voltage noise spectra $S_u(f)$ are different because the temperatures of the resistors are different at practical (non-ideal) conditions. However, in practical cases, the voltage noise generators are not thermal but external generators [2] with 1-100 million times greater voltages than the thermal noise therefore the temperature has no relevance and Hao's comments are inappropriate.

In Section 2, we show several reasons why Feng Hao's comments [1] are irrelevant. In Section 3, we show how these comments can be modified and be made partially relevant for practical situations within the class of non-ideality problems of security. Finally, we show the solution how to keep the practical security at any required level, beyond quantum security. This is the very same method that was already described in the first paper [2] and security data were published in [6].

## 2. Why are Feng Hao's comments irrelevant?

### 2.1. *Irrelevance for the case of idealized (totally secure) KLJN communicators*

It is important to emphasize that the comments [1] are irrelevant for the idealized/mathematical KLJN key exchange scheme [2] because the comments [1] violate the mathematical conditions assumed for total security. In the KLJN scheme it is assumed that the temperatures or the spectra of voltage noise generators are equal and violating that key assumption naturally leads to compromised security. It is like assuming that quantum communicators requiring single photons for security cannot produce single photons but only larger packages of photons, which is a valid practical problem at quantum communication. Though all the practical quantum communicators suffer from this deficiency compromising practical security, we do not say that the idealized/mathematical quantum communicator schemes are insecure. Therefore, we conclude that *Feng Hao's comments violate basic assumptions thus they are irrelevant for the theoretical KLJN scheme which remains totally secure*.

### 2.2. *Irrelevance for the case of practical KLJN communicators*

From the very beginning [2], *Johnson (-like) noise* was used for the KLJN scheme and the Johnson noise based considerations were used only for educational purpose, for the sake of simplicity, because the well known statistical physical characteristics of Johnson noise make the total security of the idealized KLJN cipher obvious.

As it has been pointed out in the very first paper [2], at practical applications, the use of thermal noise is disadvantageous (except stealth communication [9]) therefore the issue of temperature and its accuracy is unimportant even for the security of common practical realizations. In practical applications, the voltage noise generators are not thermal but external voltage noise generators [2] with 1-100 million times greater RMS noise voltage than the Johnson noise. *Therefore the temperature has no relevance and Feng Hao's comments are inappropriate for the practical KLJN realizations, too*.

## 3. A valid practical case: inaccuracies as non-idealities and how to treat them

It is important to emphasize that any deviation from the idealized scheme of a physical secure communication scheme can lead to a finite level of security. Relevant examples for that are practical quantum communicators with the impossibility of generating strictly ideal single photon packages or with the impossibility to avoid detector noise which also compromise the quantum security. Thus, if the above-mentioned author had stressed the inaccuracy of the resistor values, his claim would have been practically valid. However the impact of such systematic errors can be kept under control in practical KLJN realizations. Here we cite our relevant analysis [6] of 1% voltage inaccuracy (voltage drop) along the line due to wire resistance effects. The results can be considered as general result for the security leak at inaccuracies of this order (1%). The necessity to build a statistics to extract information during the properly chosen clock time leads to extremely poor eavesdropper statistics and this situation provides an excellent practical *raw bit security* that can easily be beyond the *raw bit security* of idealized quantum communicators.

Figure 2 shows the results of an analysis of the information leak due to systematic errors/inaccuracies [6], with 1% relative effect, which is a significantly worse situation than the 0.5% systematic temperature error supposed by Feng Hao [1]. In the case of Feng Hao's supposed situation, Eve would need 4 times longer clock period to extract the same amount of information as the analysis indicates below. Practical parameters and noise rectifiers have been assumed (see [6]). The noise bandwidth (correlation time) and clock time period are selected so that Alice, Bob and Eve are able to extract only a few independent samples during the clock period. Alice and Bob must decide between the amplitude distributions $f$ and $g$ on Figure 2 (a) and Eve must decide between the amplitude distributions $f$ and $g$ on Figure 2 (b) while they have a few random hits at the x coordinates. It is obvious that Alice and Bob have an easy decision and Eve has a very poor case. Still, the application of Shannon's channel code theorem indicates that, if Eve finds out the best possible decoding method, she may be able to extract about 0.7% of the exchanged bits. However this raw bit leak is still better than raw bit leaks in idealized quantum communicators (see [6]).

*Response to Feng Hao's comments about security versus temperature inaccuracies.*

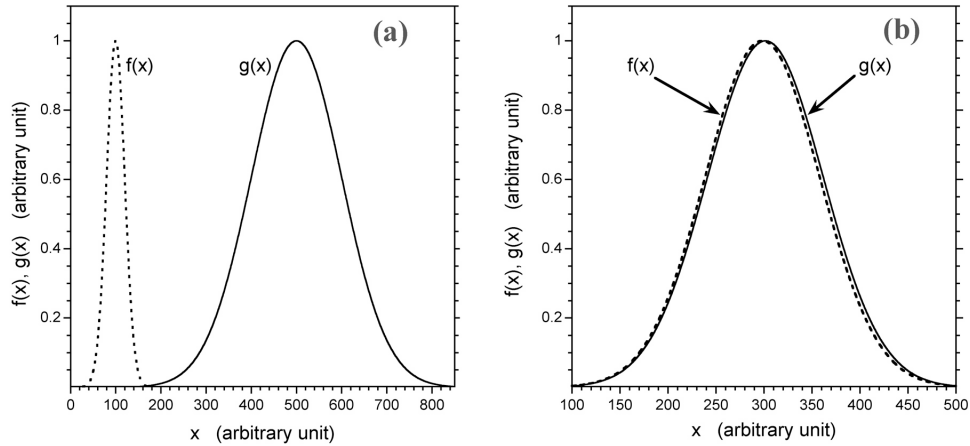

**Figure 2.** Analysis of the information leak due to systematic errors/inaccuracies [6], with 1% relative effect, which is a significantly worse situation than the 0.5% systematic error supposed by Hao. In the case of Hao's situation, Eve would need a 4 times longer clock time period to reach the same situation as indicated by these amplitude density functions.

Finally we note that 1% inaccuracy is an "economical" one and, if higher raw bit security is needed, the accuracy can be improved with sufficient resources. For example, a 10 times increase of the wire diameter decreases the 1% inaccuracy (voltage drop) in [6] to 0.01% which would make the curves *f* and *g* in Figure 2 (b) indistinguishable by the naked eye and decrease the bit leak to 0.00007% which is a *raw bit leak* /security much beyond the reach of any quantum communicator.